\documentclass[twocolumn,showpacs,prl,amsmath,amssymb]{revtex4}

\usepackage{graphicx}% Include figure files
\usepackage{dcolumn}% Align table columns on decimal point
\usepackage{bm}% bold math
\usepackage{amsmath}
\usepackage{amssymb}
\usepackage{mathrsfs}
\usepackage{color}
\begin{document}
% \draft
\title{Quantum spirals}
% \title{Generating Vorticity in Quantum Fluids}
\author{Zensho Yoshida$^1$ and Swadesh M. Mahajan$^2$}
\affiliation{$^1$Graduate School of Frontier Sciences, The University of Tokyo,
Kashiwa, Chiba 277-8561, Japan
\\
$^2$Institute for Fusion Studies, The University of Texas at Austin, Austin,Texas 78712, USA}
\date{\today}

\begin{abstract}
Quantum systems often exhibit fundamental incapability to entertain {vortex}.
The Meissner effect, a complete expulsion of the magnetic field  (the electromagnetic vorticity), for instance, 
is taken to be the defining attribute of the superconducting state.
Superfluidity is another, close-parallel example; 
fluid vorticity can reside only on topological defects with a limited (quantized) amount.
Recent developments in the Bose-Einstein condensates produced by particle traps further emphasize
this characteristic. 
We show that the challenge of imparting vorticity to a quantum fluid can be met through 
a nonlinear mechanism operating in a hot fluid corresponding to a thermally modified 
Pauli-Schr\"odinger  spinor field. 
In a simple field-free model, we show that the thermal effect, 
represented by a nonlinear, non-Hermitian Hamiltonian, in conjunction with spin vorticity, leads to 
new interesting quantum states; a spiral solution is explicitly worked out.  
\end{abstract}

\pacs{ 03.75.Kk, 67.10.Hk, 52.35.We, 47.10.Df}
% 52.35.We Plasma vorticity
% 52.72.+v Laboratory studies of space&astrophysical-plasma processes 
%   52.35.Fp Electrostatic waves and oscillations
% 45.20.Jj Lagrangian and Hamiltonian mechanics  
% 47.10.Df	fluid--Hamiltonian formulations
% 67.10.Hk	Quantum effects on the structure and dynamics of non-degenerate fluids
% 03.75.Kk	Dynamic properties of condensates; collective and hydrodynamic excitations, superfluid flow
\maketitle

\section{Introduction}
\label{sec:introduction}

The \emph{magic} of quantum correlations has been often invoked to explain the exotic and
extraordinary phenomena like the superfluidity, superconductivity, and Bose-Einstein condensation, 
displayed by, what may be called, the \emph{quantum fluids}\,\cite{London-I, London-II,Landau, Gross, Pitaevskii, Hohenberg, Fischer,Fetter}.
What links all these diverse systems is the absence of vorticity. 
The Meissner effect, a complete expulsion of the magnetic field  (the electromagnetic vorticity), for instance, 
is taken to be the defining attribute of the superconducting state\,\cite{London-I,Landau}
(see also \cite{mahcs} for a characterization of superconductivity as vanishing of the total vorticity, a sum of the fluid and electromagnetic vorticities).

Of course, these highly correlated quantum states are accessible only under very special conditions, in particular at very low temperatures. 
It is, perhaps, legitimate to infer that these quantum fluids, when they are not in there \emph{super} phase, may, in fact, 
entertain some sort of vorticity. 
In this article, we explore how such a vortical state may emerge in a quantum  system, 
whose basic dynamical equations (like the Schr\"odinger equation)
% (because of their Hamilton-Jacoby structure) 
are not fundamentally suitable for hosting a vortex.

The investigation of the quantum vortex states constitutes a fundamental enquiry,
because such states, like their classical counterparts, aught to be ubiquitous. 
And again as to their classical counterparts, the vorticity will lend enormous diversity and complexity
to the behavior and properties of quantum fluids. 
Unveiling the mechanisms responsible for creating/sustaining vorticity
will, surely, advance our understanding of the dynamics of the {phase transition} 
---from zero vorticity to finite vorticity and vice versa.

We will carry out this deeper enquiry, aimed at bridging the \emph{vorticity gap}, by exploring  
a quantum system equivalent to a hot fluid/plasma\,\cite{mahase2014}. 
We will show that the thermodynamic forces induce two distinct fundamental changes in the dynamics: 
the Hamiltonian becomes 1) nonlinear by the thermal energy, and 2) non-Hermitian by the entropy.
By these new effects, a finite vorticity becomes accessible to the quantum fluid.
%  characteristic "wave function" obeys
% a nonlinear equation, and 2) finite vorticity becomes accessible to the quantum fluid.
Such a vorticity-carrying hot quantum fluid could define a new and interesting state of matter.
Within the framework of a hot Pauli-Schr\"odinger quantum fluid, we will
demonstrate the existence of one such state---the Quantum Spiral. 
We believe that it marks the beginning of a new line of research.

% To set the stage for our construction, 
To highlight the new aspect of our construction,
we present a short overview of papers on the classical-quantum interplay.
% This paper draws heavily from two sets of original papers on the subject of classical-quantum interplay; 
The first set of investigations\,\cite{mad,bohm,tak1,tak2,tak3,tak4,cufaro,  Fro,haas,marklund1,marklund2,vortical,asenjorqp} 
is devoted to deriving (and studying) the fluid-like systems from the standard equations of quantum mechanics, 
while the second set\,\cite{kania,pesci1,pesci2,Andreev, carbonaro, Koide, Kambe}
constructs a quantum mechanics equivalent to a given classical system.
% starting with either  a classical Hamiltonian \cite{pesci1,pesci2} or an energy momentum tensor \cite {kania, mahase2014}.
% An important lesson we draw from these works is that a quantum vorticity is fundamentally connected to the spin. 
Building from the energy momentum tensor for a perfect isotropic hot fluid, 
Mahajan \& Asenjo\,\cite {mahase2014} have recently demonstrated 
that the emergent quantum mechanics of an elementary constituent of the Pauli-Schr\"odinger hot fluid (called a \emph{fluidon}) 
is nonlinear as distinct from the standard linear quantum mechanics; the thermal interactions manifest as the fluidon self-interaction.
Through a deeper reexamination of this {thermal interaction},
% (via the Hamiltonian formalism of a generic nonlinear Pauli-Schr\"odinger spinor field) 
we begin our quest for the thermal mechanism of creating quantum vorticity.
In addition to being a source of nonlinearity, the thermal interaction for the quantum fluid 
also endows it with an {entropy}.
% to complete the story, we must and we have identified the entropy with an appropriate function of the  wave function.  

\section{Circulation law ---the Vortex and Heat}
Theres exists a deep relationship between the vortex and a heat cycle,
which is mediated by an entropy.
We may consider a \emph{vortex} in general space;
by {vortex} we mean finiteness of a \emph{circulation} (or, non-exact differential).
A {heat cycle} $\oint T d{S} \neq 0$ ($T$ is the temperature and ${S}$ is the entropy)
epitomizes such a circulation.

Upon the realization of the thermodynamic law on a fluid, the heat cycle is related to the circulation of momentum,
i.e., the mechanical vortex.
For a fluid with a density $\rho$, pressure ${P} $, and enthalpy $H$, 
obeying the thermodynamic relation $Td{S}=d{H} - \rho^{-1} d{P} $,
a finite heat cycle $\oint T d{S} \neq 0$ is equivalent to the \emph{baroclinic effect} $\oint \rho^{-1} d{P} \neq 0$
(the exact differential $dH$ does not contribute a circulation; $\oint dH\equiv 0$).
Notice that the entropy, a deep independent attribute of thermodynamics,
is the source of the baroclinic effect; such an effect will be encountered whenever a
field has a similar internal degree of freedom represented by some scalar like an entropy.
% Notice that the entropy is the deep structure underlying this thermodynamic effect.
% When we consider a field that has an internal degree of freedom represented by some scalar like an entropy,
% we encounter a `baroclinic effect'.
 
Kelvin's circulation theorem says that, as far as the specific pressure force $\rho^{-1} d{P} $ 
(or, the heat $Td{S}$) is an exact differential,
the fluid (so-called barotropic fluid) conserves the circulation $\oint_\Gamma \wp $ of the momentum $\wp$
along an arbitrary co-moving loop $\Gamma$.
Therefore, a vorticity-free flow remains so forever.
As the antithesis, non-exact thermodynamic force $\rho^{-1} d{P} $
(or, equivalently, a heat cycle $\oint_\Gamma Td{S}\neq0$)
violates the conservation of circulation of momentum,
leading to the vortex creation.

% Consider a fluid with a density $\rho$, temperature$T$, pressure ${P} $, enthalpy $H$, entropy ${S}$
%  obeying the thermodynamic relation $Td{S}=d{H} - \rho^{-1} d{P} $. 
% Let $\wp$ denote a momentum 1-form.
% The fluid is said to ``host a vortex'' if the \emph{circulation} $\oint_\Gamma \wp$ is finite
% ($\Gamma$ is an arbitrary loop). 
% In a typical dynamics, a finite circulation is guaranteed when $\oint T d{S} \neq 0$ (or equivalent $\oint \rho^{-1} d{P} \neq 0$), i.e, 
%  when either of these combinations is not an exact differential. This statement is the  antitheses of the 
%  Kelvin circulation theorem that says that, as far as the specific pressure force $\rho^{-1} d{P} $ is an exact differential, 
% the fluid conserves the circulation along an arbitrary co-moving loop $\Gamma$. The non exact thermodynamic force,
% leading to the vortex creation (violation of the conservation of circulation)is the cause of the so called \emph{ baroclinic effect}.
%  

%%%%%%%%%%%%%%%%%%%%%%%%%%%%%%%%%%%%%
\section{Vortex in spinor fields}
To formulate a quantum-mechanical baroclinic effect by quantizing a classical fluid model,
the {correspondence principle} is best described by the Madelung representation of wave functions\,\cite{mad,bohm} (see Appendix A).
We must, however, remember that a scalar (zero-spin) Schr\"odinger field falls short of describing a vortex,
because the momentum field is the gradient of the eikonal of the Schr\"odinger field
(or, in the language of classical mechanics, the momentum is the gradient of the action),
which is evidently curl-free.
This simple fact is, indeed, what prevents a conventional  Schr\"odinger quantum regime from hosting vorticity.

In a set of classic papers~\cite{tak1,tak2,tak3,tak4}, Takabayasi showed that 
the differences in the phases of spinor components could 
generate a \emph{spin vorticity} in the `Madelung fluid' equivalent of
the Pauli-Schr\"odinger quantum field. 
In this paper, we investigate 
an additional source of vorticity provided by a baroclinic mechanism in a thermally 
modified nonlinear Pauli-Schr\"odinger system, obtained here, by adding a thermal energy $U$
to the Pauli-Schr\"odinger Hamiltonian. 

In order to delineate the new effect in the simplest way, we consider a minimum, field-free Hamiltonian 
\begin{equation}
\mathscr{H} = \int \left[ \frac{1}{2m} \left(\frac{\hbar}{i}\nabla\Psi\right)^* \cdot
\left(\frac{\hbar}{i}\nabla\Psi\right)
+ {U} \Psi^*\cdot \Psi \right]
\,dx ,
\label{Q-Hamiltonian-spinor}
\end{equation}
where $\Psi=(\psi_1, \psi_2)$ is a two-component spinor field,
% (may be regarded as a spin 1/2 representation of SU(2), or the zero-trace conjugacy group of SO(3)),
and $U$ is a thermal energy.
Notice that the conventional potential energy is replaced by the thermal energy.
% Here we assume no magnetic field, and ignore the gauge symmetry
% (which can be easily incorporated by replacing $\nabla$ by a covariant operator).
The formulation of ${U}$ as a function of $\Psi$ is the most essential element of our construction.
In classical thermo/fluid dynamics, the thermal energy is generally expressed as $U=U(\rho,S)$
with the density $\rho$ and the entropy $S$.
Although $\rho$ is readily expressed as $\Psi^*\Psi$, finding an expression for $S$ is more challenging. 
It is the right juncture to inform the reader that for $U=U(\rho)$, the sought after baroclinic effect is absent; see \cite{mahase2014}. 

The Madelung representation of the wave function
\begin{equation}
\psi_j (\bm{x},t) = \sqrt{\rho_j(\bm{x},t)} e^{i\mathscr{S}_j(\bm{x},t)/\hbar}
\quad (j=1,2) .
\label{Madelung-2}
\end{equation}
converts the two complex field variables $(\psi_1, \psi_2)$ into four real variables $(\rho_1,\mathscr{S}_1,\rho_2,\mathscr{S}_2)$.
It will be, however, more convenient to work with an equivalent set:
% (we call them the \emph{Clebsch-Takabayasi variables}):
\begin{equation}
\left\{ \begin{array}{l}
\rho = \rho_1+\rho_2, \\%= \Psi^*\Psi\\,
\mu= \rho_1-\rho_2, \\
\varphi = (\mathscr{S}_1+\mathscr{S}_2)/2, \\
\sigma = (\mathscr{S}_1-\mathscr{S}_2)/2.
\end{array} \right.
\label{Clebsch}
\end{equation}
The 4-momentum becomes
\begin{equation}
{p}^\nu = \frac{1}{\rho} \Re \left( \Psi^*\cdot i\hbar \partial^\nu \Psi  \right)
% &=& -\frac{1}{\rho} \left(\rho_1 \partial^\mu\mathscr{S}_1 + \rho_2 \partial^\mu \mathscr{S}_2 \right)
% \nonumber
% \\
= -\left( \partial^\nu \varphi + \frac{\mu}{\rho}\partial^\nu \sigma \right),
\label{momentum}
\end{equation}
where  $(x_0,x_1,x_2,x_3)=(t,-x,-y,-z)$,
% $(x_0,x_1,x_2,x_3)=(t,-x,-y,-z)$, $(\partial_0,\partial_1,\partial_2,\partial_3)=(\partial_t,\nabla)$,
and $(\partial^0,\partial^1,\partial^2,\partial^3)=(\partial_t,-\nabla)$).
The spatial part of (\ref{momentum}),
\begin{equation}
\bm{p}  = \nabla\varphi +  \frac{\mu}{\rho} \nabla \sigma
\label{Clebsch2}
\end{equation}
reads as the Clebsch-parameterized momentum field\,\cite{Clebsch,Lin,Jackiw,Yoshida_Clebsch}.
The second term of the right-hand side of (\ref{Clebsch2}) yields a
\emph{vorticity}:
\begin{equation}
\nabla\times\bm{p}  =\nabla \left( \frac{\mu}{\rho}\right) \times \nabla \sigma .
% = -\frac{\hbar}{2} (\nabla {S}_1\times\nabla {S}_2)/{S}_3.
\label{spin_vorticity2''}
\end{equation}
In (\ref{spin_vorticity2''}), we have assumed that $\varphi$ does not have a phase singularity.
If $\varphi$ is an angular (multi-valued) field, 
as in the example of quantum spirals given later, 
a circulation, representing a \emph{point vortex}, will be created by the singularity of $\nabla\times(\nabla\varphi)$ 
(mathematically, a cohomology). 

We may regard $\sigma$ as a Lagrangian label of scalar fields co-moving with the fluid
(see Appendix A).
Hence, for an isentropic process,
we parameterize entropy as ${S}=S(\sigma)$ to put $U=U(\rho,\sigma)$, completing 
the process of identification of the thermal variables with the wave function.
The enthalpy and temperature are, respectively, given by
\begin{equation}
{{H}} = \frac{\partial(\rho {U} )}{\partial \rho},
\quad 
{{T}} = \frac{\partial {U}}{\partial{S}}  .
\label{enthalpy}
\end{equation}
Denoting $S'(\sigma)=dS(\sigma)/d\sigma$, we may define an effective temperature
$\tau=\partial U/\partial \sigma = S'(\sigma) T$.

\section{Thermally-modified nonlinear Pauli-Schr\"odinger equation}
We are ready to derive the determining equation.
In terms of $\Psi$, the canonical 1-form reads
$\Theta = \int p^0 \rho \,dx$.
The variation of the action $\int (\Theta - \mathscr{H})\,dt$ by $\Psi$
yields a thermally-modified nonlinear Pauli-Schr\"odinger equation:
\begin{equation}
i\hbar\partial_t \psi_j =
 - \frac{\hbar^2}{2m} \nabla^2\psi_j + \left( {{H}} -  i{{G}}_j \right) \psi_j
\quad (j=1,2) ,
\label{Schroedinger-baroclinic}
\end{equation}
where 
\begin{equation}
 {{G}}_j = (-1)^j \hbar \frac{ {{S'(\sigma)T}} \rho}{4 \rho_j} \quad (j=1,2).
\label{baroclinic_term}
\end{equation}

The following results can be readily derived by the new equation (\ref{Schroedinger-baroclinic}):

% Observe that in the the new equation (\ref{Schroedinger-baroclinic}):
% \begin{enumerate}
% \item
(i) The terms $-iG_j \psi_j$ ($j=1,2$) on the right-hand side of (\ref{Schroedinger-baroclinic}) 
represent the {baroclinic effect}, by which the generator of the system is {non-Hermitian}.
However, the the particle number ($\int\rho\,dx$) and the energy ($\mathscr{H}$) are preserved
as constants of motion.

\begin{figure}[tb]
\raisebox{3.2cm}{\textbf{a}}
\includegraphics[scale=0.6]{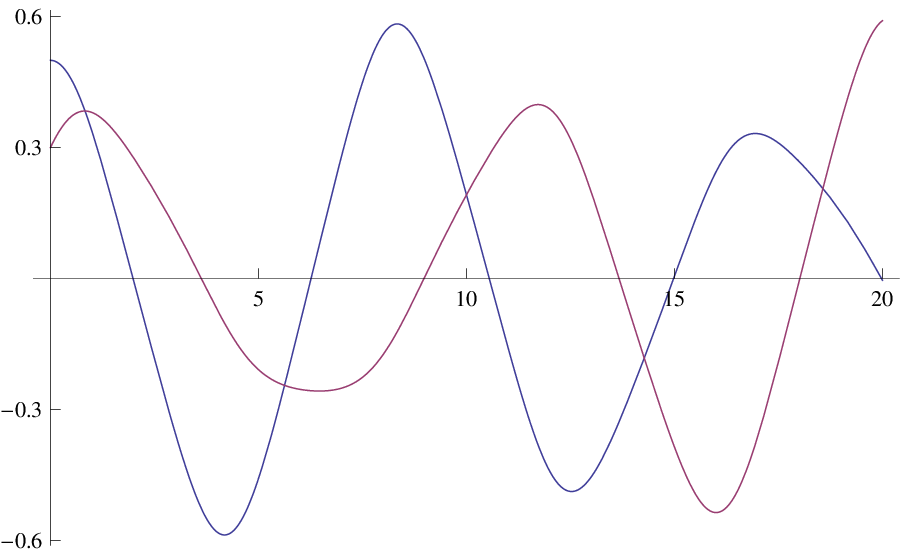}
\\ ~ \\
\raisebox{3.2cm}{\textbf{b}}
\includegraphics[scale=0.6]{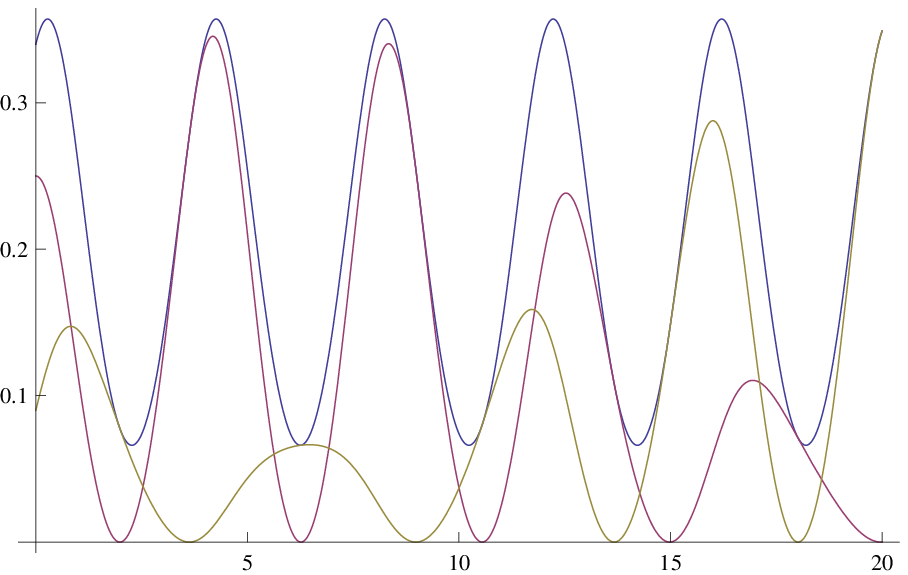}
\caption{
\label{fig:1D}
A typical nonlinear Pauli-Schr\"odinger field in one-dimensional space (here the eigenvalue $\mu=0$).
(\textbf{a}) The two spinor components
(here, real-valued functions)  $\psi_1$ (blue) and $\psi_2$ (red) are coupled though the nonlinear enthalpy coefficient
$H = -2(|\psi_1|^2+|\psi_2|^2)$, exhibiting chaotic oscillations.
(\textbf{b}) The densities $\rho_1 = |\psi_1|^2$ (red) and $\rho_2 = |\psi_2|^2$ (yellow) oscillates irregularly,
while the total density $\rho = \rho_1 + \rho_2$ (blue) remains ordered.
}
\end{figure}

% \item
(ii) When $S'(\sigma)=0$ (i.e., the fluid is homentropic), the baroclinic terms are zero.
Then, (\ref{Schroedinger-baroclinic}) consists of two
coupled nonlinear Schr\"odinger equations (the nonlinear coupling comes through $H$ being a function of $\rho$).
Of course, we may put $\psi_2\equiv 0$, and then, the system reduces into the standard scalar-field nonlinear Schr\"odinger equation
governing $\psi_1$.
It is well known that, in a one-dimensional space, we obtain \emph{solitons}, when ${{H}}=a \rho$ ($a<0$).
The nonlinear coupling of the two components $\psi_1$ and $\psi_2$ induces chaotic behavior.
Interestingly, however,  $\rho = |\psi_1|^2 + |\psi_2|^2$ remains ordered. 
These features are displayed in Fig.\,\ref{fig:1D}, where a representative solution in the barotropic ($G_j=0$) limit is plotted.

%---------------------------------------------------------------  FIG 1
%----------------------------------------------------------------------

% \item
(iii) When the baroclinic terms are finite,
there is no one-dimensional (plane wave) solution.
In fact, upon substitution of $\psi_j=e^{i(k_yy+k_zz-\omega t)/\hbar}\phi_j(x)$ ($j=1,2$) into (\ref{Schroedinger-baroclinic}), 
we obtain an eigenvalue problem
 \begin{equation}
\frac{\hbar^2}{2m} \frac{d^2}{dx^2} \phi_j = (\lambda + {{H}} -i {{G}}_j ) \phi_j
\quad (j=1,2),
\label{Schroedinger-baroclinic-1D}
\end{equation}
where $\lambda = (k_y^2+k_z^2)/2m -\omega$. Half of the (local, i.e., for each $\psi_j$)
eigenvalues for this operator, 
$\pm \sqrt{\lambda+ {{H}} -i {{G}}_j}$, have
positive real parts whenever ${{G}}_j\neq 0$,
and thus, (\ref{Schroedinger-baroclinic-1D}) cannot have a bounded solution for any $\lambda$.

% \end{enumerate}

The nonexistence of a one-dimensional (plane-wave) solution in a baroclinic system emphasizes the fact that the 
baroclinic effect is absent in a one-dimensional system. 
However, we  do find interesting solutions in two-dimensional space.

\section{Quantum spirals}
Let us assume a solution of a spiral form:
\begin{equation}
\Psi =
\left(\begin{array}{c}
\psi_1 \\ \psi_2 \end{array} \right)
=
\left(\begin{array}{cc}
 e^{i(n\theta + \beta_1(r) -\omega t)} \phi_1(r)
 \\
 e^{i(n\theta + \beta_2(r) -\omega t)} \phi_2(r)
\end{array} \right) .
\label{spiral-1}
\end{equation} 
The azimuthal mode number $n$ gives the number of arms.
The phase factor $\beta_j(r)$ determines their shape;
for example, when $\beta_j(r)$ is a linear function of $r$, we obtain a Archimedean spiral.
The factor $\phi_j(r)$ yields the radial modulation of amplitudes.
We find that the nonlinear terms
\begin{eqnarray}
\rho &=& % \rho_1+\rho_2= \psi_1^*\psi_1+\psi_2^*\psi_2
\phi_1^* \phi_1+\phi_2^* \phi_2 ,
\label{spiral-2}
\\
\sigma &=& % \frac{1}{2}( \mathscr{S}_1 - \mathscr{S}_2) =
% \frac{\hbar}{2} (\beta_1 - \beta_2 ) - i \frac{\hbar}{4} \left[ \log(\phi_1/\phi_1^*) - \log(\phi_2/\phi_2^*) \right]
\frac{\hbar}{2} \left( \beta_1 - \beta_2  + \arg \phi_1 - \arg \phi_2 \right)
\label{spiral-3}
\end{eqnarray}
are functions only of $r$; the azimuthal mode number $n$, therefore, is a good quantum number.
Inserting (\ref{spiral-1}) into (\ref{Schroedinger-baroclinic}), we obtain ($j=1, 2$, $\dot{~}$=$d/dr$),
\begin{eqnarray}
\ddot{\phi}_j + \left(\frac{1}{r} + i\dot{\beta}_j\right) \dot{\phi}_j  =
\left[\left( -\omega + \frac{n^2}{r^2} + \dot{\beta}_j^2 
+  {{H}} \right) 
 \right.
\nonumber
\\
\left.
- i\left(\ddot{\beta}_j +\frac{\dot{\beta}_j}{r}-  {{G}}_j\right) \right] \phi_j .
% \quad (j=1,2).
\label{Schroedinger-baroclinic-spiral}
\end{eqnarray}
Bounded solutions are obtained when the non-Hermitian term vanishes; one must, then, solve  
the system of simultaneous equations
\begin{eqnarray}
\ddot{\phi}_j + \left(\frac{1}{r} + i\dot{\beta}_j\right) \dot{\phi}_j  &=&
\left( -\omega + \frac{n^2}{r^2} + \dot{\beta}_j^2 +  {{H}} \right)  \phi_j ,
\label{Schroedinger-baroclinic-spiral-1}
\\
\ddot{\beta}_j + \frac{\dot{\beta}_j}{r} - {{G}}_j &=&  0.
% \quad (j=1,2).
\label{Schroedinger-baroclinic-spiral-2}
\end{eqnarray}
By (\ref{Schroedinger-baroclinic-spiral-2}), the baroclinic term $G_j$
generates $\beta_j(r)$ that determines the shape of spiral.
Evidently, in a barotropic field ($G_j=0$), spirals do not appear ($\beta_j(r)=0$).

To construct explicit examples,
let us consider an \emph{ideal gas} that has an internal energy such that
\begin{equation}
{U} = c_v (\rho e^{{S}-{\sigma}_0} )^{1/c_v},
\label{ideal_gas}
\end{equation}
where $c_v$ (specific heat normalized by Rydberg constant) and ${\sigma}_0$ are constants.
For simplicity, we assume $S=\sigma$ (thus, $T=\tau$)
to evaluate 
\begin{equation}
{{T}} = (\rho e^{{\sigma}-{\sigma}_0} )^{1/c_v},
\quad
{{H}}=(c_v +1) {{T}}.
\label{ideal_gas-2}
\end{equation}
Substituting (\ref{spiral-2}) and (\ref{spiral-3}), we may write
the coefficients $H$ and $G_j$ in terms of $\phi_j$ and $\beta_j$ ($j=1,2$).
% We assume ${S} = ({S}'/\hbar) \sigma$ to write
% \begin{equation}
% e^{\sigma} = e^{\hbar[(\beta_1-\beta_2) + (\arg \phi_1 - \arg \phi_2)]/2}.
% \label{ideal_gas-3}
% \end{equation}

\begin{figure}[tb]
\includegraphics[scale=0.3]{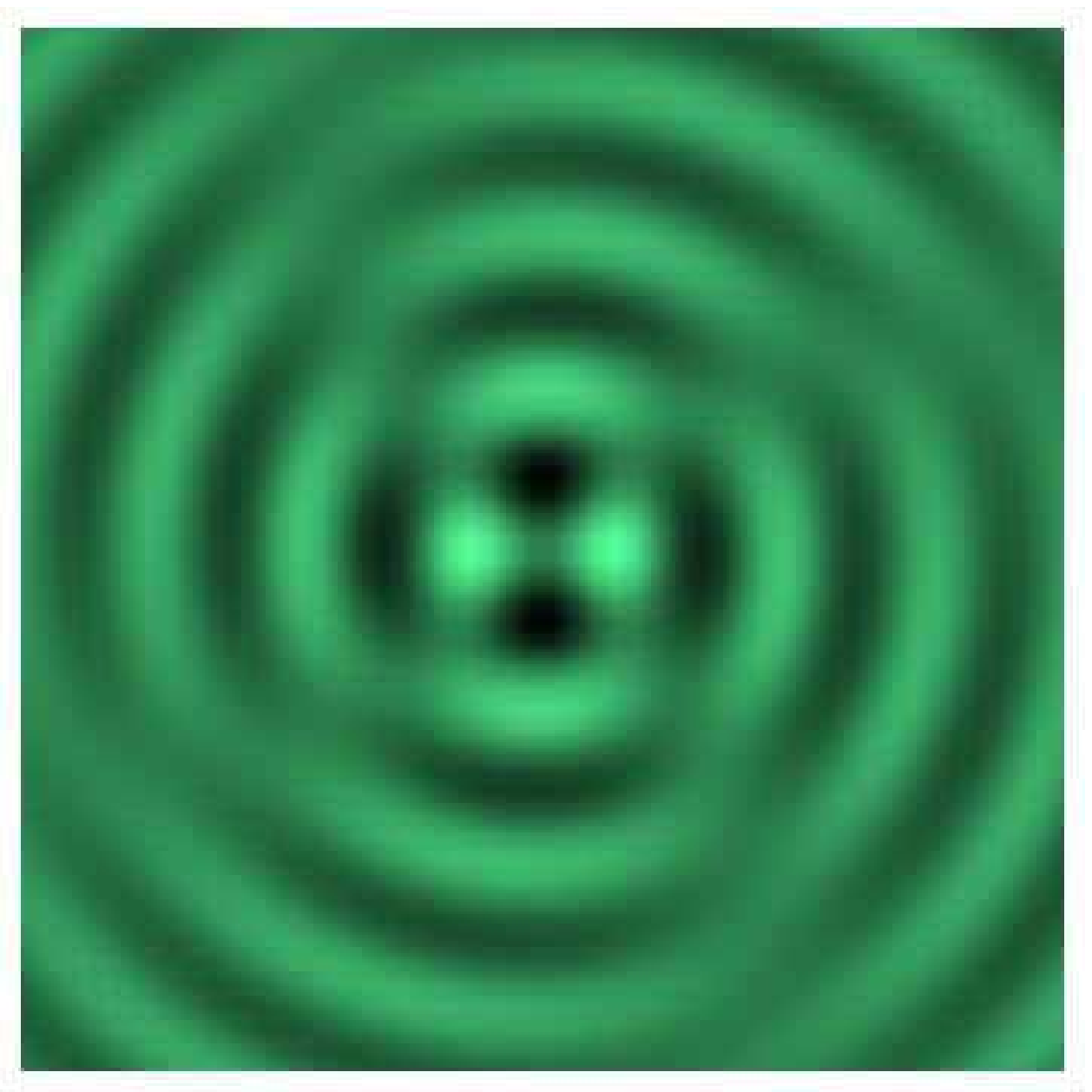}
~~~
\includegraphics[scale=0.3]{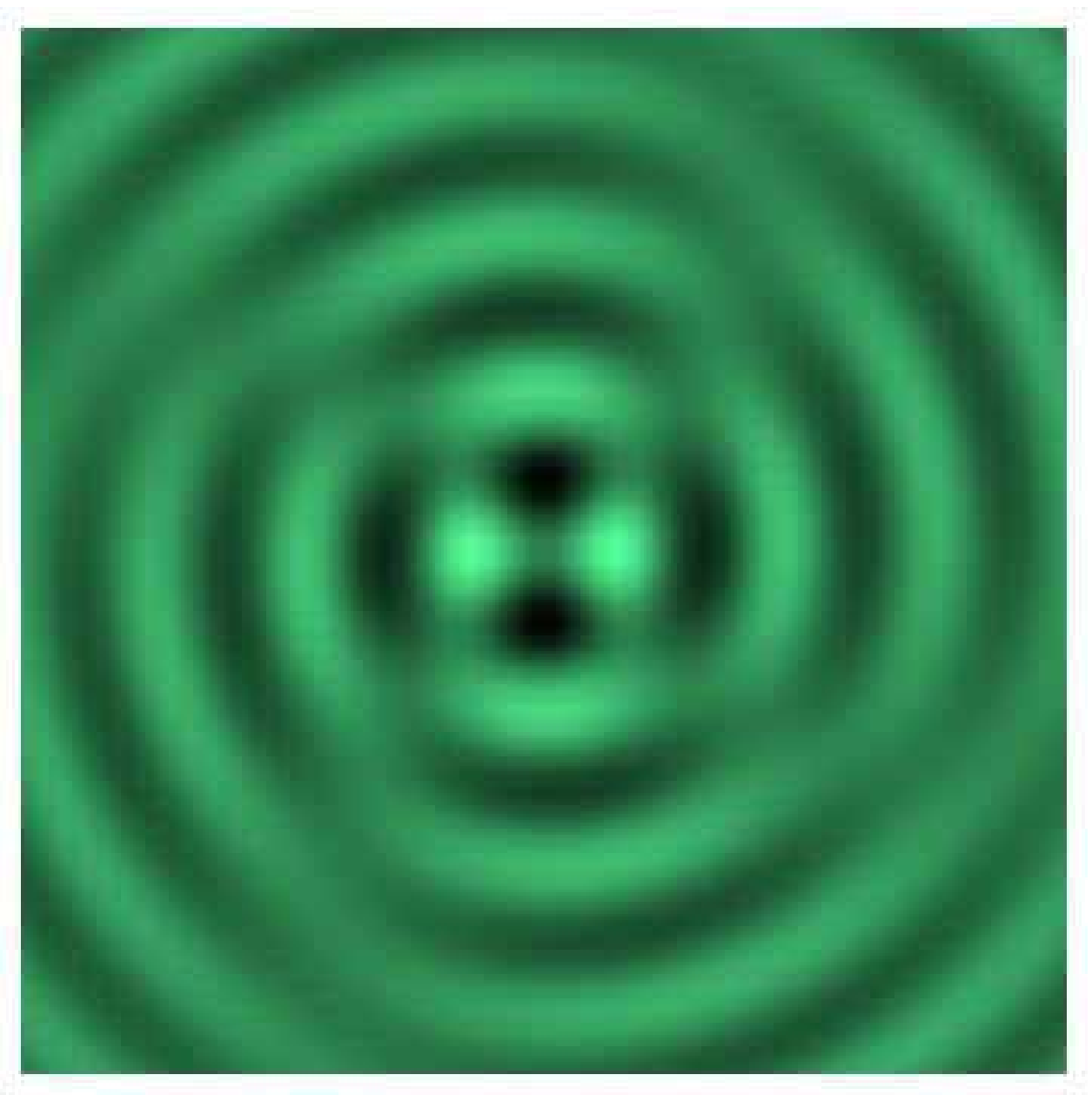}
\caption{
\label{fig:baroclinic_m2}
Dual spirals created in a baroclinic Pauli-Schr\"odinger field;
the density plots of $\Re \psi_1$ (left) and $\Re \psi_2$ (right)
show opposite-sense spirals.  
Here $\omega=4.5$.
}
\end{figure}

\begin{figure}[tb]
\begin{center}
\raisebox{3.2cm}{\textbf{a}}~
\includegraphics[scale=0.6]{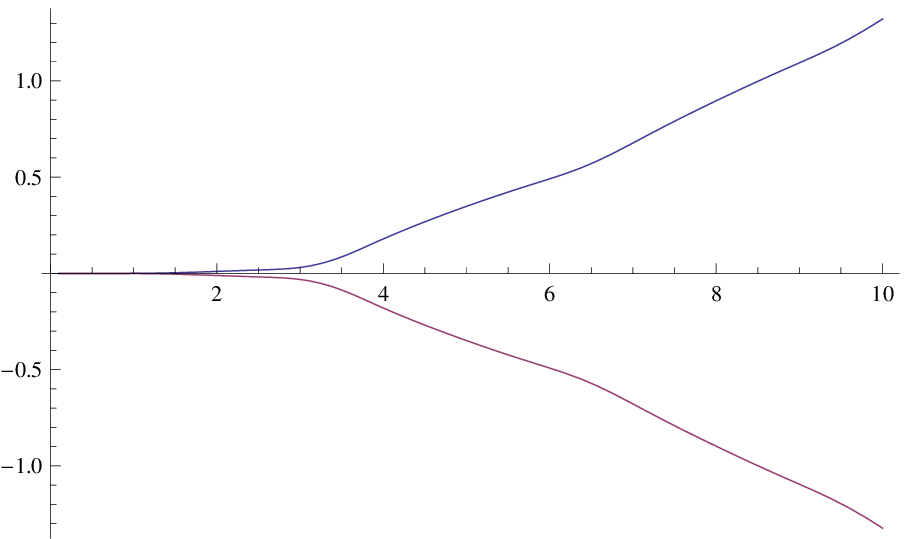}
\\ ~ \\
\raisebox{3.2cm}{\textbf{b}}~
\includegraphics[scale=0.6]{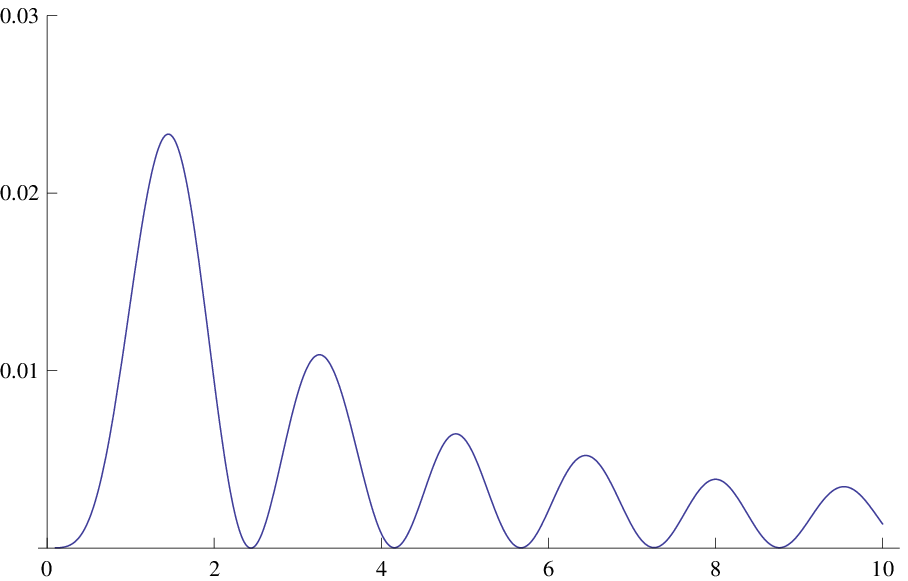}
\caption{
\label{fig:baroclinic_m2_fields}
(\textbf{a}) The phase factors $\beta_1(r)$ (blue) and $\beta_2(r)$ (red)
of the quantum spirals shown in Fig.\,\ref{fig:baroclinic_m2}.
Except in the core region ($r<3$), $\beta_j(r)$ is an approximately linear function of $r$,
hence the arms are approximately Archimedean.
The opposite signs of $\beta_1(r)$ and $\beta_2(r)$ give opposite-sense spirals.
(\textbf{b})  The amplitude factors $|\phi_1(r)|^2$ ($=|\phi_2(r)|^2$ by the assumption).
}
\end{center}
\end{figure}

As an example ($c_V=1$ and $\sigma_0=0$) of the numerical solutions of this system,
we display in Fig.\,\ref{fig:baroclinic_m2}, a typical $n=2$ solution exhibiting twin spirals (opposite sense) of
the two components of the spinor $\Psi$.
Figure\,\ref{fig:baroclinic_m2_fields}
picks up the phase factors $\beta_j(r)$ and the amplitude factors $|\phi_j(r)|^2$
from the spiral spinor fields of Fig.\,\ref{fig:baroclinic_m2}.

%---------------------------------------------------------------  FIG 2
%----------------------------------------------------------------------

%---------------------------------------------------------------  FIG 3
%----------------------------------------------------------------------

\begin{figure}[tb]
\begin{center}
\raisebox{3.5cm}{\textbf{a}}
\includegraphics[scale=0.3]{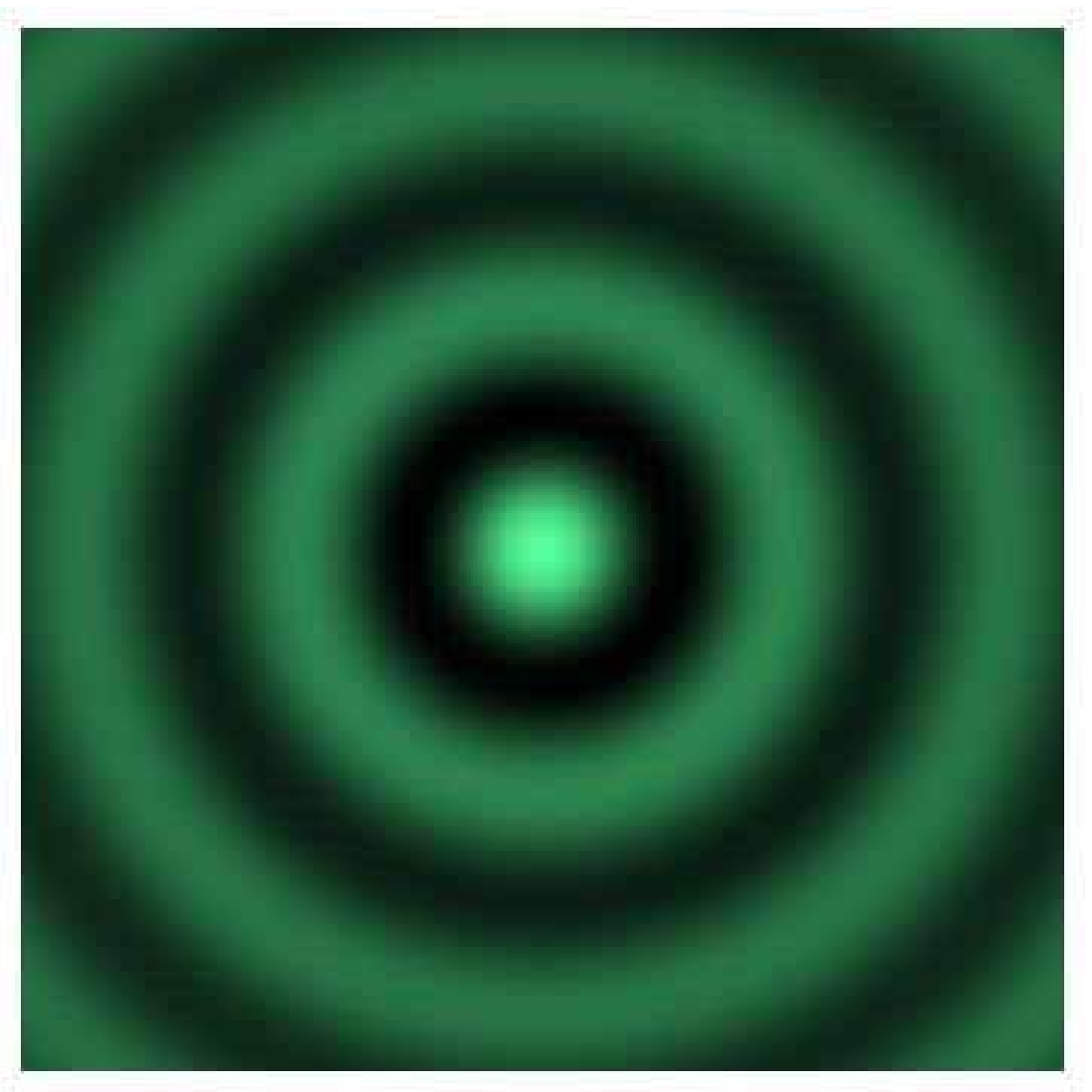}
~~
\raisebox{3.5cm}{\textbf{b}}
\includegraphics[scale=0.3]{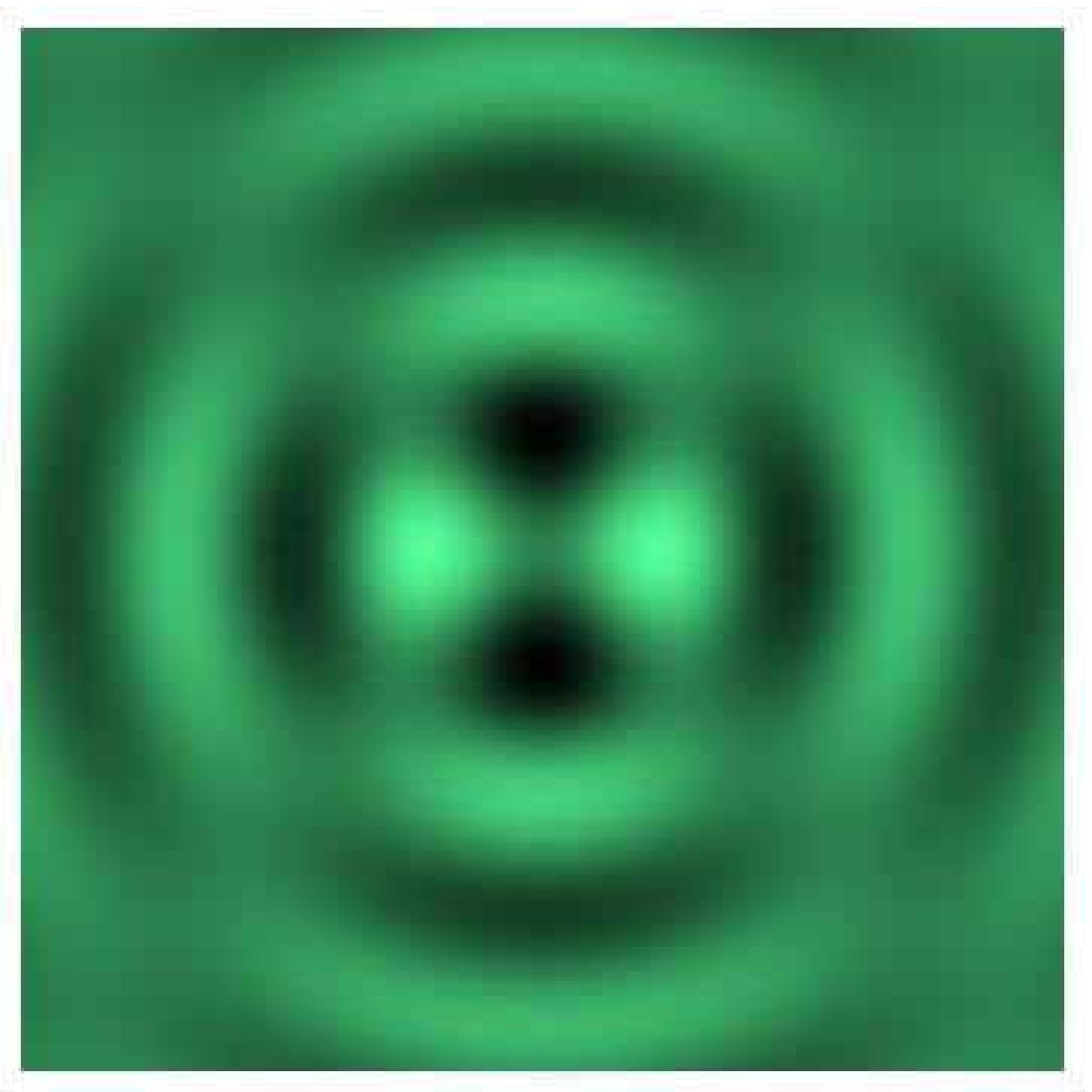}
\caption{
\label{fig:baroclinic_m0}
(\textbf{a}) Axisymmetric ($n=0$) solution of the baroclinic equation with $\omega=4.5$;
plot of $\Re \psi_1$.
(\textbf{b}) Barotropic ($G_j=0$) Pauli-Schr\"odinger field ($n=2$) with $\omega=4.5$;
plot of $\Re \psi_1$.
When $G_j=0$, the phase factor $\beta_j(r)=0$, thus no arms are created.
}
\end{center}
\end{figure}

The baroclinic  Pauli-Schr\"odinger equation has axisymmetric ($n=0$) solutions also (see Fig.\,\ref{fig:baroclinic_m0}-a).

Figure\,\ref{fig:baroclinic_m0}-b shows that when the baroclinic term is zero 
(then, the generator is Hermitian),
no spiral structures, even with a finite $n$, are created.
This is because the phase factors $\beta_j(r)$ become zero when $G_j=0$; see (\ref{Schroedinger-baroclinic-spiral-2}).

%---------------------------------------------------------------  FIG 4
%----------------------------------------------------------------------

\section{Conclusion}
We have shown that the quantum fluid with internal thermal energy is capable of supporting 
entirely new quantum states like the quantum spirals we have derived in the present simplified model.
The harnessing of the profound effect of entropy, in a thermal spin quantum system, has
led us to a new mechanism 
(whose classical counterpart is the famous baroclinic effect generating,
for example, hurricanes) 
that substantially extends the range of vortical states accessible to quantum systems.
 
We note that although the conventional spin forces 
% It must be appreciated, however, that  although  the conventional spin forces 
(either spin-magnetic field interactions or spin-spin interaction; cf. Appendix A)
may amplify or sustain vorticities, they cannot generate it from zero.
The baroclinic effect is a creation mechanism that works without a \emph{seed}.
A close analogy is the respective roles of dynamo amplification of magnetic field\,\cite{Moffatt} 
and the Biermann battery mechanism\,\cite{Biermann} in a classical plasma; the former needs a seed magnetic field.   

We also note that the baroclinic effect we have studied is an isentropic process,
and differs from dissipation mechanisms like friction.
For example, a finite-temperature Bose gas is modeled by a modified Gross-Pitaevskii equation 
including an effective friction term, which is coupled with
a quantum Boltzmann equation describing a thermal cloud\,\cite{ZNG,Jackson,Allen}
(see also tutorial\,\cite{Proukakis}).
The friction term introduces a non-Hermitian Hamiltonian which, in contrast to the presently
formulated isentropic model, destroys the conservation of the particle number and  the energy of the condensate component.

One expects to find a variety of new/interesting states when this work is extended
for more encompassing quantum systems: the thermal Pauli-Schr\"odinger system and  the relativistic thermal Dirac (Feynman-Gellmann)
equation coupled to the electromagnetic field.
For both of these systems, baroclinic terms can be readily incorporated in 
known formalisms \cite{vortical,Koide,mahase2014}. 
Coriolis force, a close cousin of the magnetic force, could also be included as a gauge field\,\cite{Kambe}.
When we consider higher-order spinors
(for example, spin-1 representation of SU(2) fields may be applied for Bose gas in a trap\,\cite{spin-1})
the number of Clebsch parameters increases, and the field starts to have a helicity\,\cite{Yoshida_Clebsch}.

%%%%%%%%%%%%%%%%%%%%%%%%%%%%%%%%%%%%%%%%%%%%%%%%%%%%%%%%%%%%%%%%%%%%%%%%%%%%%%%%%%%%%%%%%%%
\section*{Acknowledgment}
A part of this work was done as a JIFT program of US-Japan Fusion Research Collaboration.
ZY's research was supported by JSPS KAKENHI Grant Number 23224014.
SMM's research was supported by the US DOE
%  Department of Energy 
Grant DE-FG02-04ER54742.

%%%%%%%%%%%%%%%%%%%%%%%%%%%%%%%%%%%%%%%%%%%%%%%%%%%%%%%%%%%%%%%%%%%%%%%%%%%%%%%%%%%%%%%%%%%
\section*{Appendix A. Correspondence principle}
We explain the correspondence principle
% ($\{ ~, ~ \} \rightarrow (i \hbar)^{-1} [~,~] $,
relating the classical and quantized fields.
In terms of the real variables (\ref{Clebsch}),  
the Hamiltonian (\ref{Q-Hamiltonian-spinor}) reads $\mathscr{H} = \mathscr{H}_c + \mathscr{H}_q $ with
\begin{eqnarray}
\mathscr{H}_c &=& \int \left[ \frac{|\nabla \varphi + \frac{\mu}{\rho}\nabla\sigma|^2}{2m} 
+ {U}(\rho,\sigma ) \right] \rho \,dx ,
\label{c-limit}
\\
\mathscr{H}_q  &=&  \int \frac{\hbar^2}{8m} \left(  \frac{|\nabla\rho|^2}{\rho^2} + \sum_\ell |\nabla {S}_\ell|^2 \right) \rho \,dx.
\label{H-quantization}
\end{eqnarray}
Here $\mathscr{H}_c$ is a classical Hamiltonian generating fluid dynamics\,\cite{Lin}.
% The first term in $\mathscr{H}_q$ is the so-called Bohm pressure.
% The spin term $\sum_\ell |\nabla {S}_\ell|^2$ is necessary to quantize the Clebsch-form (finite vorticity) momentum
% into a spinor.
% Although $\mathscr{H}_q$ is multiplied by the factor $\hbar^2$, 
% it includes order-unity terms stemming from the derivatives $\nabla {S}_\ell$
% (since ${S}_\ell$ includes the phase factor $e^{i\mathscr{S}_j/\hbar}$).
% This is in marked contrast to the case of scalar wave function.
% The classical equation of motion is derived by the classical Hamiltonian $\mathscr{H}_c$.
With the canonical 1-form
% \begin{equation}
$\Theta  = \int {p}^0 \rho dx = -\int (\rho \partial_t\varphi + \mu \partial_t\sigma )\,dx $,
% \label{canonical_1-form-Clebsch}
% \end{equation}
the variation of the classical action $\int (\Theta - \mathscr{H}_c)\,dt$
with respect to the canonical variables $(\rho,\varphi,\mu,\sigma)$ yields Hamilton's equation:
\begin{eqnarray}
\partial_t{\rho} &=& \partial_{\varphi} \mathscr{H}_c = -\nabla\cdot(\bm{v}\rho) ,
\label{fluid_canonical-1}
\\
\partial_t{\varphi} &=& -\partial_\rho \mathscr{H}_c = -\bm{v}\cdot\nabla\varphi+m|\bm{v}|^2/2- H,
\label{fluid_canonical-2}
\\
\partial_t{\mu} &=& \partial_{\sigma} \mathscr{H}_c = -\nabla\cdot(\bm{v}\mu) + \rho T,
\label{fluid_canonical-3}
\\
\partial_t{\sigma} &=& -\partial_{\mu} \mathscr{H}_c = -\bm{v}\cdot\nabla\sigma ,
\label{fluid_canonical-4}
\end{eqnarray}
where $\bm{v} =\bm{p}/m$ is the fluid velocity,
% These equations describe classical ideal fluid mechanics; 
(\ref{fluid_canonical-1}) is the mass conservation law, 
(\ref{fluid_canonical-4}) is the entropy conservation law
(justifying the parameterization $U(\rho,\sigma)$),
and the combination of all equations with the thermodynamic relation $\nabla {{H}} - {{T}} \nabla{\sigma} = \rho^{-1} d {P} $ ($P$ is the pressure) yields the momentum equation
\begin{equation}
% & & \partial_t\rho + \nabla\cdot(\rho\bm{v}) =0 ,
% \label{fluid_canonical-5}
% \\
\partial_t\bm{p} + (\bm{v}\cdot\nabla)\bm{p} = - \rho^{-1}\nabla {P}.
\label{fluid_canonical-6}
\end{equation}
The system (\ref{fluid_canonical-1})-(\ref{fluid_canonical-4}) is an infinite-dimensional 
Hamiltonian system endowed with a canonical Poisson bracket such that
\begin{equation}
\{ \rho(\bm{x}), \varphi(\bm{y}) \} = \delta(\bm{x}-\bm{y}), 
~
\{ \mu(\bm{x}), \sigma(\bm{y}) \} = \delta(\bm{x}-\bm{y});
% \\
% \{ \rho(\bm{x}),  \rho(\bm{y}) \} =0, \{ \varphi(\bm{x}),  \varphi(\bm{y}) \} =0,
% \quad
% \{ \varphi(\bm{x}),  \varphi(\bm{y}) \} =0,
\label{canonical_bracket-1}
\end{equation}
all other brackets are zero.
By (\ref{Madelung-2}) and (\ref{Clebsch}),
the Poisson algebra (\ref{canonical_bracket-1})
is equivalent to the Lie algebra of the second-quantized Pauli-Schr\"odinger field
acting on the Fock space of either Bosons or Fermions\,\cite{Jackiw}:
\begin{equation}
\begin{array}{l}
{[} \psi_j(\bm{x}), \psi_k^*(\bm{y}) {]}_\pm = (i\hbar)^{-1} \delta_{jk} \delta(\bm{x}-\bm{y}),
\\ 
{[} \psi_j(\bm{x}), \psi_k(\bm{y}) {]}_\pm  =0,
~
{[} \psi_j^*(\bm{x}), \psi_k^*(\bm{y}) {]}_\pm =0.
\end{array}
\label{canonical_bracket-0}
\end{equation}
Based on this correspondence principle, we can quantize the classical field by adding $\mathscr{H}_q$ to $\mathscr{H}_c$;
the action principle with respect  to $\Psi$ yields (\ref{Schroedinger-baroclinic}).
The same action principle with respect to the real variables (\ref{Clebsch}) yields the fluid representation:
on the right-hand side of (\ref{fluid_canonical-6}), $\mathcal{H}_q$ adds quantum forces\,\cite{tak1,tak2}
\begin{equation}
\bm{F}_q=
\nabla \left( \hbar^2 \frac{\nabla^2\sqrt{\rho}}{2m\sqrt{\rho}}\right)
-\sum_\ell\frac{\hbar^2}{4m\rho}\nabla\cdot[\rho\nabla {S}_\ell\otimes \nabla {S}_\ell] ,
\label{fluid_canonical-7} 
\end{equation}
where
${S}_\ell =\rho^{-1}\Psi^* \bm{\sigma}_\ell \Psi $ 
($\bm{\sigma}_\ell$ are the Pauli matrices). %\,\cite{SO3}.

% While the first term (Bohm pressure) is curl-free, the second term (spin force) can produce vorticity.
% (we note that $\nabla {S}_\ell$ yields a term of order $\hbar^{-1}$; hence,
% the spin force includes an order-unity contribution).
% However, it is finite only when a seed vorticity exists.
% In contrast, the baroclinic drive ($-\rho^{-1}\nabla P$, which already exists in the classical regime)
% can work even if the initial vortex is zero, i.e., $\mu=\rho_1-\rho_2=0$
% (which may be compared with the Biermann battery mechanism\,\cite{Biermann} versus dynamo amplification of magnetic field\,\cite{Moffatt} in a classical plasma; the latter needs a seed magnetic field).

%%%%%%%%%%%%%%%%%%%%%%%%%%%%%%%%%%%%%%%%%%%%%%%%%%%%%%%%%%%%%%%%%%%%%%%%%%%%%%%%%%%%%%%%%%%

%%%%%%%%%%%%%%%%%%%%%%%%%%%%%%%%%%%%%%%%%%%%%%%%%%%%%%%%%%%%%%%%%%%%%%%%%%%%%%%%%%%%%%%%%%%
%%%%%%%%%%%%%%%%%%%%%%%%%%%%%%%%%%%%%%%%%%%%%%%%%%%%%%%%%%%%%%%%%%%%%%%%%%%%%%%%%%%%%%%%%%%
%%%%%%%%%%%%%%%%%%%%%%%%%%%%%%%%%%%%%%%%%%%%%%%%%%%%%%%%%%%%%%%%%%%%%%%%%%%%%%%%%%%%%%%%%%%
%%%%%%%%%%%%%%%%%%%%%%%%%%%%%%%%%%%%%%%%%%%%%%%%%%%%%%%%%%%%%%%%%%%%%%%%%%%%%%%%%%%%%%%%%%%
%%%%%%%%%%%%%%%%%%%%%%%%%%%%%%%%%%%%%%%%%%%%%%%%%%%%%%%%%%%%%%%%%%%%%%%%%%%%%%%%%%%%%%%%%%%


\begin{thebibliography}{99}

\bibitem{London-I}
London, F.
\textit{Superfluids, Vol. I: Macroscopic Theory of Superconductivity.}
(Wiley, 1950).
% (Wiley, New York,1950).

\bibitem{London-II}
London, F.
\textit{Superfluids, Vol. II: Macroscopic Theory of Superfluid Helium.}
(Wiley, 1954).

\bibitem{Landau}
Landau, L. \& Lifshitz, E.
\textit{Statistical Physics}, Chap. VI 
(Addison-Wesley, 1958).
% (Addison-Wesley, Reading, 1958).

% \bibitem{Onsager}
% L. Onsager, Nuovo Cimento, Suppl. \textbf{6}, 249 (1949).

% \bibitem{Feynman}
% R. P. Feynman, Prog. Low Temp. Phys. \textbf{1}, 17 (1955).

\bibitem{Gross}
Gross, E. P.
Structure of a Quantized Vortex in Boson Systems.
\textit{Nuovo Cimento} \textbf{20}, 454--477 (1961).

\bibitem{Pitaevskii}
Pitaevskii, L. P. 
Vortex Lines in an Imperfect Bose Gas,
\textit{Sov. Phys. JETP} \textbf{13}, 451--454 (1961).

\bibitem{Hohenberg}
Hohenberg, P. C. \& Martin, P. C. 
Microscopic theory of superfluid helium,
\textit{Ann. Phys.} \textbf{34}, 291--359 (1965).


\bibitem{Fischer}
Fischer, U. R. \& Baym, G. 
Vortex states of rapidly rotating dilute Bose-Einstein condensates,
\textit{Phys. Rev. Lett}. \textbf{90}, 140402 (2003).

\bibitem{Fetter}
Fetter, A. L. 
Rotating trapped Bose-Einstein condensates,
\textit{Rev. Mod. Phys.} \textbf{81}, 647--691 (2009).

\bibitem{mahcs}
Mahajan, S. M. 
Classical Perfect Diamagnetism: Expulsion of Current from the Plasma Interior,
\textit{Phys. Rev. Lett.} \textbf{100}, 075001 (2008).

\bibitem{mahase2014} 
Mahajan, S. M. \& Asenjo, F. A. 
Hot Fluids and Nonlinear Quantum Mechanics,
\textit{Int J Theor Phys.} \textbf{54}, 1435--1449 (2015).


\bibitem{mad} 
Madelung, E. 
Quantentheorie in hydrodynamischer Form,
\textit{Z. Phys.} {\bf 40}, 322--326 (1927).

\bibitem{bohm} 
Bohm, D. 
A Suggested Interpretation of the Quantum Theory in Terms of ``Hidden'' Variables. I,
\textit{Phys. Rev.} {\bf 85}, 166--178 (1952).

\bibitem{tak1} 
Takabayasi, T. 
On the Hydrodynamical Representation of Non-Relativistic Spinor Equation,
\textit{Prog. Theor. Phys.} {\bf 12}, 810--812 (1954).

\bibitem{tak2} 
Takabayasi, T. 
The Vector Representation of Spinning Particle in the Quantum Theory I,
\textit{Prog. Theor. Phys.} {\bf 14}, 283--302 (1955).
% \bibitem{tak3} T. Takabayasi, 
% Variational Principle in the Hydrodynamical Formulation of the Dirac Field,
% Phys. Rev. {\bf 102}, 297--298 (1956).

\bibitem{tak3} 
Takabayasi, T. 
Hydrodynamical description of the Dirac equation,
\textit{Nuovo Cimento} {\bf 3}, 233--241 (1956).
% \bibitem{tak5} T. Takabayasi, 
% Relativistic Hydrodynamics Equivalent to the Dirac Equation,
% {\bf 13}, 222--224 (1955); 

\bibitem{tak4} 
Takabayasi, T. 
Vortex, Spin and Triad for Quantum Mechanics of Spinning Particle I,
\textit{Prog. Theor. Phys.} {\bf 70}, 1--17 (1983); 
% \\
% Prog. Theor. Phys. Suppl. 4, 2 (1957).

\bibitem{cufaro} 
Cufaro Petroni, N., Gueret, Ph. \& Vigier, J.-P. 
A causal stochastic theory of spin-1/2 fields,
\textit{Nuovo Cimento B} {\bf 81}, 243--259 (1984).

% \bibitem{aron} 
% J. Aron,  ???
% Compt. Rend. {\bf 251}, 921 (1960) 

\bibitem{Fro}
Fr\"ohlich, H. 
Microscopic derivation of the equations of hydrodynamics,
\textit{Physica A} {\bf 37}, 215--226 (1967).

\bibitem{haas} 
Haas, F. 
\textit{Quantum plasmas: An Hydrodynamic Approach} 
(Springer, 2011).

\bibitem{marklund1} 
Marklund, M. \& Brodin, G. 
Dynamics of Spin-1/2 Quantum Plasmas,
\textit{Phys. Rev. Lett.} {\bf 98}, 025001 (2007).

\bibitem{marklund2} 
Brodin, G. \& Marklund, M. 
Spin magnetohydrodynamics,
\textit{New J. Phys.} {\bf 9}, 277 (2007).

\bibitem{vortical} 
Mahajan, S. M. \& Asenjo, F. A.
Vortical Dynamics of Spinning Quantum Plasmas: Helicity Conservation,
\textit{Phys. Rev. Lett.} {\bf 107}, 195003 (2011).

\bibitem{asenjorqp} 
Asenjo, F. A., Mu\~noz, V., Valdivia, J. A. \& Mahajan, S. M. 
A hydrodynamical model for relativistic spin quantum plasmas,
\textit{Phys. Plasmas} {\bf 18}, 012107 (2011).

\bibitem{kania} 
Kaniadakis, G. 
Statistical origin of quantum mechanics,
\textit{Physica A} {\bf 307}, 172--184 (2002).

\bibitem{pesci1} 
Pesci, A. I. \& Goldstein, R. E. 
Mapping of the classical kinetic balance equations onto the Schrodinger equation,
\textit{Nonlinearity} {\bf 18}, 211--226 (2005).

\bibitem{pesci2} 
Pesci, A. I., Goldstein, R. E. \& Uys, H. 
Mapping of the classical kinetic balance equations onto the Pauli equation,
\textit{Nonlinearity} {\bf 18}, 227--235 (2005).

% A. I. Pesci, R. E. Goldstein and H. Uys,
% Mapping of the relativistic kinetic balance equations onto the Klein-Gordon and second-order Dirac equations,
% Nonlinearity {\bf 18}, 1295--1304 (2005).

\bibitem{Andreev}
Andreev, P. A. \& Kuz'menkov, L. S.
Problem with the single-particle description and the spectra of intrinsic modes of degenerate boson-fermion systems,
\textit{Phys. Rev. A} \textbf{78}, 053624 (2008).

\bibitem{carbonaro} 
Carbonaro, P. 
\textit{EPL} {\bf 100}, 65001 (2012).

\bibitem{Koide}
Koide, T. 
Spin-electromagnetic hydrodynamics and magnetization induced by spin-magnetic interaction,
\textit{Phys. Rev. C} \textbf{87}, 034902 (2013).

\bibitem{Kambe}
Kambe, T. 
On the rotational state of a Bose-Einstein condensate,
\textit{Fluid Dyn. Res.} \textbf{46}, 031418 (2014). 

%   DOI 10.1007/s10773-014-2341-0 (2014).

\bibitem{Clebsch}
Clebsch, A. 
Uber die Integration der hydrodynamischen Gleichungen,
\textit{J. Reine Angew. Math.} \textbf{56}, 1--10 (1859).

\bibitem{Lin}
Lin, C. C.
Hydrodynamics of Helium II
in \textit{Proc. Int. Sch. Phys. ``Enrico Fermi'' XXI} 
93--146
(Academic Press, 1963).
% C. C. Lin, in \textit{Proc. Int. School of Physics `Enrico Fermi' XXI} (Academic Press, New York, 1963) p 93.

\bibitem{Jackiw}
Jackiw, R. 
\textit{Lectures on fluid dynamics ---a particle theorist's view of supersymmetic, non-Abelian, noncommutative fluid mechanics and d-branes} 
(Springer, 2002).

\bibitem{Yoshida_Clebsch}
Yoshida, Z.
Clebsch parameterization: basic properties and remarks on its applications,
\textit{J. Math. Phys.} {\bf 50}, 113101 (2009). 

\bibitem{spin-1}
Ho, T.-L. 
Spinor Bose Condensates in Optical Traps,
\textit{Phys. Rev. Lett.} \textbf{81}, 742--745 (1998). 

\bibitem{Moffatt}
Moffatt, H. K. 
\textit{Magnetic Field Generation in Electrically Conducting Fluids} 
(Cambridge University Press, 1978).

\bibitem{Biermann}
Biermann, L. 
\"Uber den Ursprung der Magnetfelder auf Sternen und im interstellaren Raum,
\textit{Z. Naturforsch.} \textbf{5a}, 65--71 (1950).

\bibitem{ZNG}
Zaremba, E. , Nikuni, T. \& Griffin, A. 
Dynamics of trapped Bose gases at finite temperatures,
\textit{J. Low Temp. Phys.} \textbf{116}, 277--345 (1999).

\bibitem{Jackson}
Jackson, B. , Proukakis, N. P. ,Barenghi, C. F. \& Zaremba, E. 
Finite-temperature vortex dynamics in Bose-Einstein condensates,
\textit{Phys. Rev. A} \textbf{79}, 053615 (2009).

\bibitem{Allen}
Allen, A. J., Zaremba, E. , Barenghi, C. F. \&Proukakis, N. P. 
Observable vortex properties in finite-temperature Bose gases,
\textit{Phys. Rev. A} \textbf{87}, 013630 (2013).

\bibitem{Proukakis}
Proukakis,N. P. \& Jackson,  B.
Finite-temperature models of Bose-Einstein condensation,
\textit{J. Phys. B: Atomic, Molecular Optical Phys.} \textbf{41}, 203002 (2008).

% \bibitem{E-M_tensor}
% The starting point for the classical to quantum transition has been  either a classical Hamiltonian \cite{pesci1} or a corresponding classical energy momentum tensor \cite {kania, mahase2014}; 
% the two approaches are equivalent.
% Building from the energy momentum tensor for a perfect isotropic hot fluid, it was demonstrated, for the first time in \cite {mahase2014}, that the emergent quantum mechanics of an elementary constituent of this hot fluid (called a fluidon) 
% is nonlinear as distinct from the standard linear quantum mechanics obtained by previous studies; 
% the thermal interactions manifest as a self-interaction of the fluidon.


% \bibitem{Morrison}
% P. J. Morrison,
% \textit{Hamiltonian description of the ideal fluid},
% {Rev. Mod. Phys.} \textbf{70}, 467 (1998).


% \bibitem{SO3}
% By the homomorphism $\textrm{SU}(2)\rightarrow\textrm{SO}(3)$
% (in the trace-zero conjugacy group), we may represent
% ${S}_1=\cos\chi \cos\delta$,
% ${S}_2=\cos\chi \sin\delta$, and
% ${S}_3=\sin\chi$.
% Comparing with (\ref{spin}), we find $\chi=\sin^{-1}(\mu/\rho)$ and $\delta=2\sigma/\hbar$.


\end{thebibliography}
\end{document}